\RequirePackage{lineno}
\documentclass[aps,prd,superscriptaddress,groupedaddress]{revtex4}
\usepackage{graphicx}
\usepackage{dcolumn}
\usepackage{bm}
\usepackage{amssymb}
\usepackage{multirow}
\usepackage{epsfig}
\usepackage{amsmath}
\usepackage{float}
\usepackage{color}
\usepackage[colorlinks, linkcolor=blue]{hyperref}

\usepackage{lineno}

\newcommand{\effstw}{\ensuremath{\sin^2\theta_{\text{eff}}^{\text{$\ell$}}}}

\newcommand{\stw}{\ensuremath{\sin^2 \theta_{W}}}

\begin{document}

\lefthyphenmin=2
\righthyphenmin=2

\widetext

\title{Reduction of the electroweak correlation in the PDF updating by using the forward-backward asymmetry of Drell-Yan process}
\affiliation{Department of Modern Physics, University of Science and Technology of China, Jinzhai Road 96, Hefei, Anhui, 230026, China}
\author{Siqi Yang}\affiliation{Department of Modern Physics, University of Science and Technology of China, Jinzhai Road 96, Hefei, Anhui, 230026, China}
\author{Yao Fu}\affiliation{Department of Modern Physics, University of Science and Technology of China, Jinzhai Road 96, Hefei, Anhui, 230026, China}
\author{Minghui Liu}\affiliation{Department of Modern Physics, University of Science and Technology of China, Jinzhai Road 96, Hefei, Anhui, 230026, China}
\author{Renyou Zhang}\affiliation{Department of Modern Physics, University of Science and Technology of China, Jinzhai Road 96, Hefei, Anhui, 230026, China}
\author{Tie-Jiun Hou}\affiliation{Department of Physics, College of Sciences, Northeastern University, Shenyang 110819, China}
\author{Chen Wang}\affiliation{Department of Modern Physics, University of Science and Technology of China, Jinzhai Road 96, Hefei, Anhui, 230026, China}
\author{Hang Yin}\affiliation{Institute of Particle Physics, Central China Normal University, Wuhan, Hubei, 430079, China}
\author{Liang Han}\affiliation{Department of Modern Physics, University of Science and Technology of China, Jinzhai Road 96, Hefei, Anhui, 230026, China} 
\author{C.--P. Yuan}\affiliation{Department of Physics and Astronomy, Michigan State University, East Lansing, MI 48823, USA}

\begin{abstract}
We propose a new observable for the measurement of the forward-backward asymmetry $(A_{FB})$ in Drell-Yan lepton production. At hadron colliders, the $A_{FB}$ 
distribution is sensitive to both the electroweak (EW) fundamental parameter $\stw$, the weak mixing angle, and the parton distribution functions (PDFs). Hence, 
the determination of $\stw$ and the updating of PDFs by directly using the same $A_{FB}$ spectrum are strongly correlated. This correlation would introduce 
large bias or uncertainty into both precise measurements of EW and PDF sectors. In this article, we show that the sensitivity of $A_{FB}$ on $\stw$ is dominated 
by its average value around the $Z$ pole region, while the shape (or gradient) of the $A_{FB}$ spectrum is insensitive to $\stw$ and contains important information 
on the PDF modeling. Accordingly, a new observable related to the gradient of the spectrum is introduced, and demonstrated to be able to significantly reduce 
the potential bias on the determination of $\stw$ when updating the PDFs using the same $A_{FB}$ data.
\end{abstract}

\maketitle
\centerline{\today}


\section{Introduction}

The parity-violating forward-backward asymmetry ($A_{FB}$) of neutral current $f_i\bar{f}_i \rightarrow Z/\gamma^* \rightarrow f_j\bar{f}_j$ fermion 
scattering processes is defined as $A_{FB} = (N_F - N_B)/(N_F + N_B)$, where $N_F$ and $N_B$ are the numbers of forward and backward events, which 
are designated according to the sign of cosine scattering angle of the outgoing fermion $f_j$ with respect to the direction of the incoming fermion $f_i$. 
In the standard model (SM), the $A_{FB}$ arises from the interference between vector and axial vector coupling terms of the $Z$ boson to fermions, 
and its value at the $Z$ boson mass pole is governed by the leptonic effective weak mixing angle parameter $\effstw$. Therefore, the $A_{FB}$ observable 
has been used to measure $\effstw$ in past decades, from the LEP and SLD experiment at electron-positron colliders to those at Tevatron and LHC hadron 
colliders, to provide precise test on the spontaneous symmetry breaking mechanism of the electroweak sector (EW) in the SM. At hadron colliders such 
as the LHC, the $A_{FB}$ spectrum is measured as a function of the invariant mass $M$ of the dilepton final states in Drell-Yan (DY) $pp\rightarrow q_i\bar{q}_i 
\rightarrow Z/\gamma^* \rightarrow l^+l^-$ processes, while the scattering angle defined in the Collins-Soper frame~\cite{CS-frame}. In the current 
framework of the SM, the quarks and antiquarks in the initial states of Drell-Yan process act as partons in the proton, and are modeled by the parton 
distribution functions (PDFs). Since quarks and antiquarks have the same probability of coming from either side of the proton beams, the observed 
asymmetry at the LHC is diluted by the uncertainty in the quark direction. This dilution effect reflects the relative difference between quarks and 
anti-quarks in proton, and thus the $A_{FB}$ at the LHC can provide unique information to constrain the PDF~\cite{xFitterPaper1, xFitterPaper2, 
ArieMassConstrain, epumpAFBsideband}, especially on the relative energy distributions of the first generation up and down quarks. 

As the $A_{FB}$ spectrum at hadron colliders is sensitive to both the PDFs and the electroweak parameter $\effstw$, extracting $\effstw$ and constraining 
PDFs by using a single $A_{FB}$ distribution are correlated, and thus would extrapolate uncertainties to each other. At the LHC experiments, the PDF-induced 
uncertainties on $\effstw$ are at $\mathcal{O}(0.1\%)$ with respect to the measured central values, presented by the ATLAS, CMS and LHCb collaborations
~\cite{ATLAS-8TeV, CMS-8TeV, LHCb-8TeV}. This PDF uncertainty would be the most dominating factor limiting the $\effstw$ measurements in the future high 
luminosity LHC experiments. On the other hand, the accuracy of updating PDF using the $A_{FB}$ at the LHC is affected by the precision of $\effstw$ 
value measured by those PDF-independent or less dependent experiments, such as the LEP/SLD~\cite{LEP-SLD} and the Tevatron~\cite{Tevatron-combine}, 
where the uncertainties on $\effstw$ are also at $\mathcal{O}(0.1\%)$ level. It has been shown~\cite{epumpAFBsideband} that the uncertainty extrapolated 
from the variation of the pre-determined $\effstw$ value into the PDF updating by using the $A_{FB}$ meausurement could be as large as the PDF-induced 
uncertainty on the $A_{FB}$ itself at the LHC. 

To take into account the above-mentioned correlation, one may argue to perform a global QCD analysis of PDFs, with $\effstw$ as an additional fitting parameter 
to the shape parameters of the non-perturbative PDFs of the proton. For example, in the CT14HERA2 PDF analysis~\cite{Dulat:2015mca, Hou:2016nqm}, there are 28 
shape parameters of the non-perturbative PDFs at the initial scale $Q_0=1.3$ GeV, from which the PDFs are evolved according to the DGLAP evolution equations. 
However, there are currently various technical difficulties to perform such an analysis in a consistent manner. First, many of the old experimental data sets 
included in the PDF global analysis were extracted at the leading order in electroweak interaction, and those data are still important for constraining different 
parton flavor ($i$) PDFs, denoted $f_i(x,Q)$, in various $x$ ranges of the proton~\cite{Hou:2019gfw}. To achieve a high precision on the determination of $\effstw$, 
all the Wilson coefficients of the scattering processes included in the PDF global analysis have to be updated to higher-order in electroweak interaction. 
This feature is not yet available in the current PDF global analysis programs. 
Second, it is known that there are tensions among various experimental data sets included in a typical, say CT14HERA2, PDF global fit. Such tensions could generate 
bias on the determination of the electroweak parameter $\effstw$ included in the fit. This is analogous to determining the strength of the coupling $\alpha_s$ in 
strong interaction via PDF global analysis. Typically, it yields a much larger uncertainty as compared to that measured from some dedicated experiments. For example,
in the recent CT18 PDF global analysis, cf. Fig.~13 of Ref.~\cite{Hou:2019efy}, it was found that the strong coupling constant, at the $Z$-boson mass ($M_Z$) scale, 
$\alpha_s(M_Z)=0.1164 \pm 0.0026$ at the $68\%$ confidence level, when including $\alpha_s$ as one of the fitting parameters of the global analysis. This result 
should be compared to the world-average value $\alpha_s(M_Z)=0.1179 \pm 0.0010$~\cite{ParticleDataGroup:2018ovx}. For that reason, the value of $\alpha_s$ is fixed in 
a typical PDF global analysis. The combined PDF+$\alpha_s$ uncertainty can be computed using the special $\alpha_s$ series of the PDFs for each family by adding 
the PDF and $\alpha_s$ uncertainties in quadrature, as explained in Ref.~\cite{Lai:2010nw}.
Third, as noted above, many non-LHC data, including Deep-Inelastic Scattering (DIS), fixed-target Drell-Yan and Tevatron data, are still important to determine 
the flavor decomposition of the PDFs in the proton, and cannot be removed all together from the PDF global analysis. Hence, it would be challenging to obtain 
a precise determination of $\effstw$ from a global QCD analysis of PDFs, with $\effstw$ as an additional fitting parameter. 
It is desirable to develop a method to independently determine the value of $\effstw$ using the precision electroweak data at the LHC. In this article, we show how 
to reduce the correlation between the PDF updating and $\effstw$ determination, by using the $A_{FB}$ spectrum from individual experiments at the future high 
luminosity LHC.

At the LHC, the $A_{FB}$-to-$\effstw$ sensitivity is dominated by the Drell-Yan events in the $Z$ boson mass pole region, with the $A_{FB}$-to-PDF sensitivity 
is higher in the off-shell sideband region~\cite{ArieMassConstrain}. Accordingly, we proposed in our previous paper~\cite{epumpAFBsideband} that 
the correlation in the PDF updating using the $A_{FB}$ spectrum can be reduced by excluding the events around the $Z$ pole. However, the potential bias raised 
by the reduced correlation derived from offshell events is still sizable. In addition, observations of the $A_{FB}$ in the sideband regions usually suffer 
large experimental difficulties and systematics. In this paper, we propose a new observable derived from the $A_{FB}$ spectrum at the LHC, which can be used 
to decouple the electroweak sectors of the Drell-Yan processes from its requirement of proton structure knowledge, and therefore build a safe experimental 
input for PDF updating and precise electroweak measurements.

\section{Definition of PDF-only-sensitive observables}

In this section, we discuss the idea of reducing the correlations to electroweak physics in the PDF updating, using the $A_{FB}(M)$ distributions at the LHC. 
All the theoretical predictions presented in this work are derived from simulations of Drell-Yan process $pp\rightarrow Z/\gamma^* \rightarrow \ell^+\ell^-$, 
generated by using ResBos~\cite{RESBOS} at 13 TeV proton-proton collision, together with the CT14HERA2 PDFs~\cite{Dulat:2015mca, Hou:2016nqm}. Samples with 
different $\effstw$ inputs are generated to demonstrate the correlation of PDFs to the EW parameter. The dilepton invariant mass region of simulation samples 
is set as $60 <M< 120$ GeV, with $80<M<100$ GeV assigned as the $Z$ pole region, and $60<M<80$ and $80<M<120$ denoted as the off-shell sideband region. 
The statistics of each sample corresponds to the size of the data collected by the ATLAS or CMS detectors during the LHC Run II period (2015-2018, 130 fb$^{-1}$).

\begin{figure}[!hbt]
\begin{center}
\includegraphics[width=0.4\textwidth]{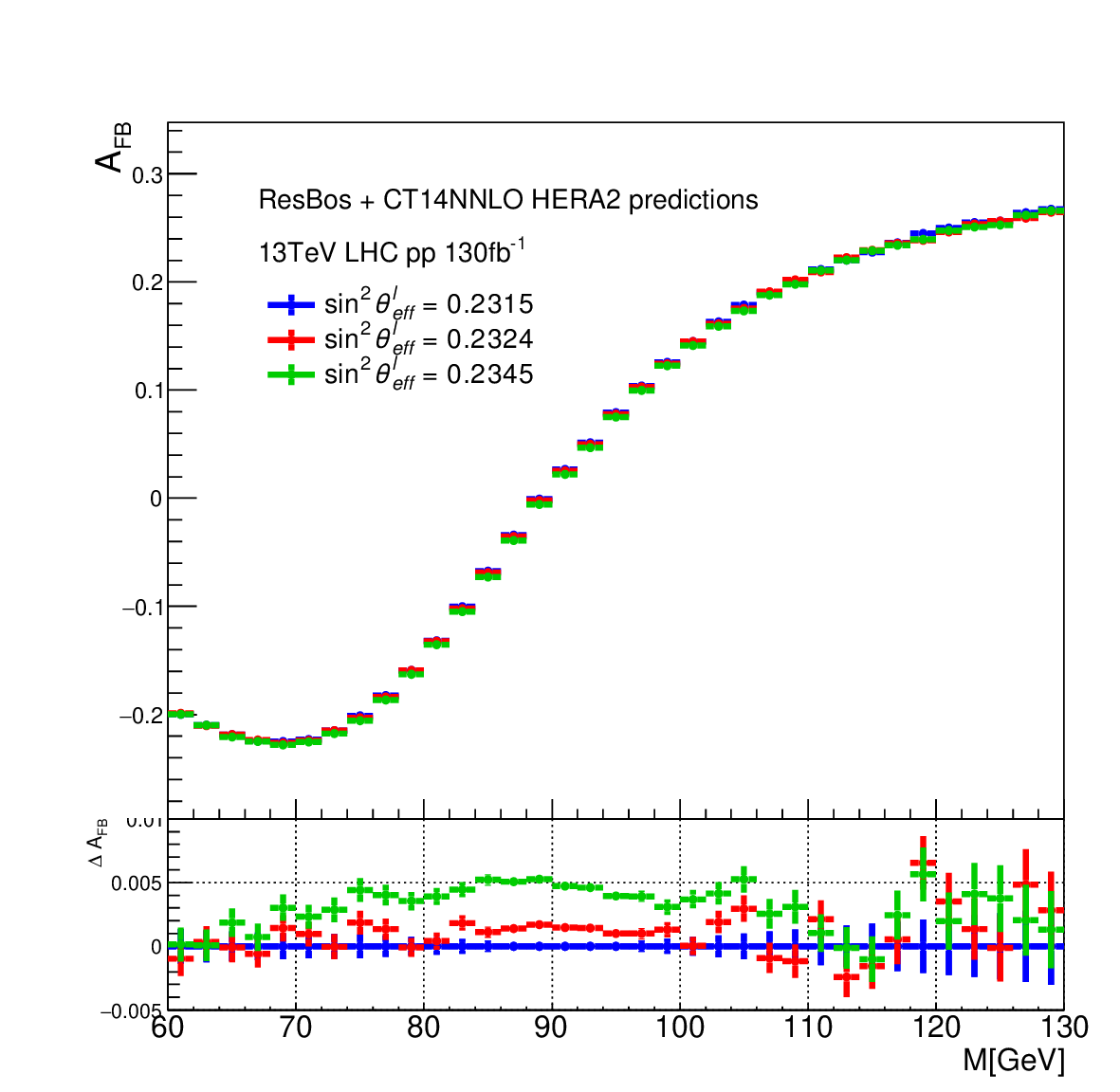}
\caption{\small Predictions of $A_{FB}(M)$ as a function of the dilepton invariant mass $M$ in $pp\rightarrow Z/\gamma^* \rightarrow \ell^+\ell^-$ events 
at 13 TeV pp collisios, with different input values of $\effstw$. The bottom panel is the bin-by-bin difference with respect to the $A_{FB}(M)$ with $\effstw=0.2315$.}
\label{fig01:AFBvsSTWatLHC}
\end{center}
\end{figure}

At hadron colliders, the scattering angle of outgoing negative charged leptons to incident quarks is calculated in the Collins-Soper frame, which is introduced 
to deal with sizable transverse momentum ($Q_T$) of the $Z$ boson and provide a kind of center of mass system of the dilepton final state observed in the laboratory 
frame.The $z$ axis in the CS frame is defined as the bisector of the angle formed by the direction of the momentum of one incoming hadron beam ($H_A$), and 
the negative direction of the momentum of the other hadron beam ($H_B$). Since the initial state of the LHC $pp$ collision is completely symmetrical, 
the $A_{FB}$ has to be measured in a modified CS frame, where $H_A$ is statistically defined as the hadron beam pointing to the same hemisphere as the reconstructed $Z$ boson pointing to. It means that the modified CS frame for the LHC is defined on event-by-event basis. Figure~\ref{fig01:AFBvsSTWatLHC} 
shows the $A_{FB}(M)$ spectrum with different input values of $\effstw$.

\begin{figure}[!hbt]
\begin{center}
\includegraphics[width=0.4\textwidth]{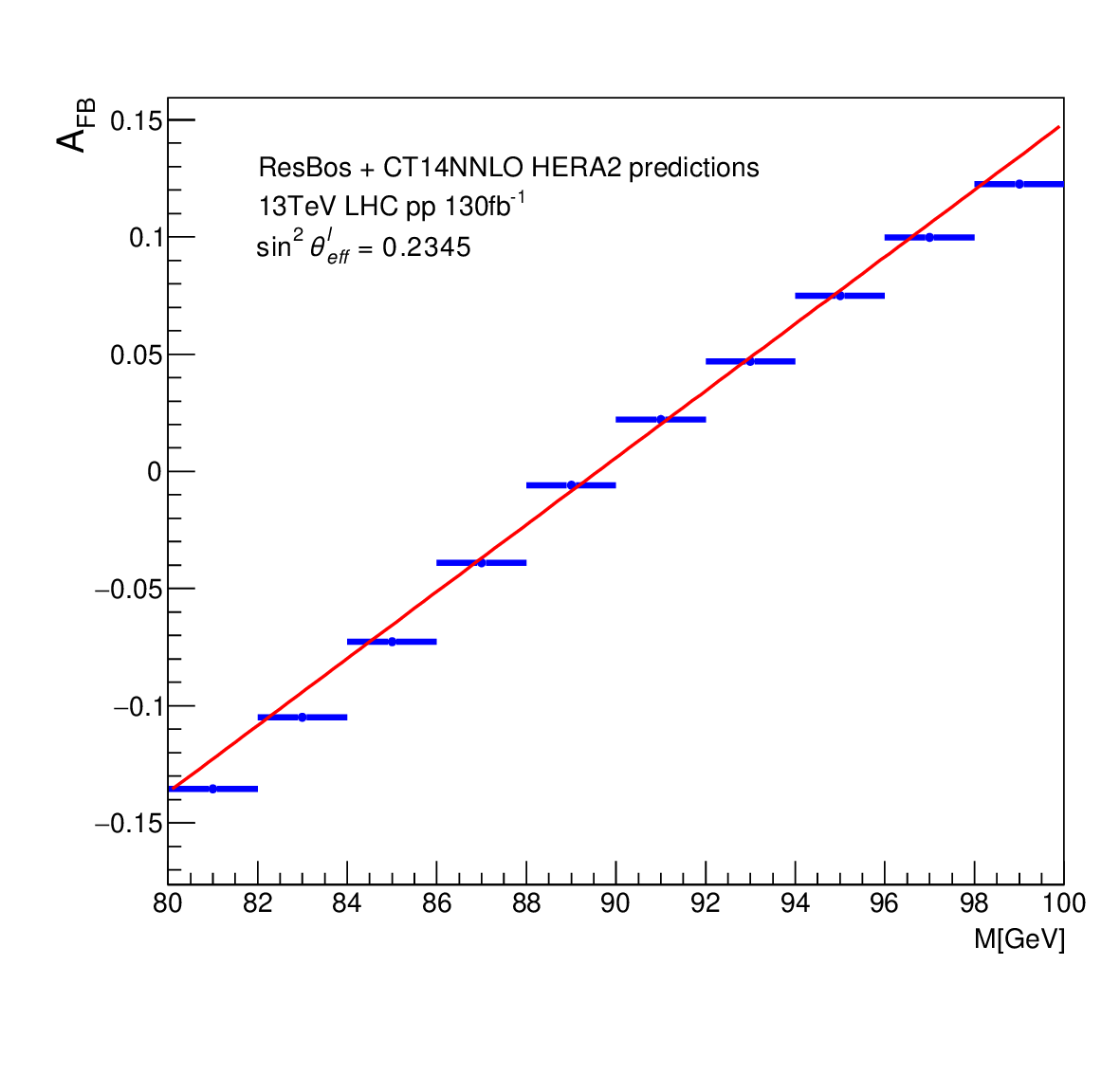}
\caption{\small Predictions of the $A_{FB}(M)$ spectrum in the $Z$-pole region at the LHC, with a linear-fitting. }
\label{fig02:AFBzpoleline}
\end{center}
\end{figure}

\begin{figure}[!hbt]
\begin{center}
\includegraphics[width=0.4\textwidth]{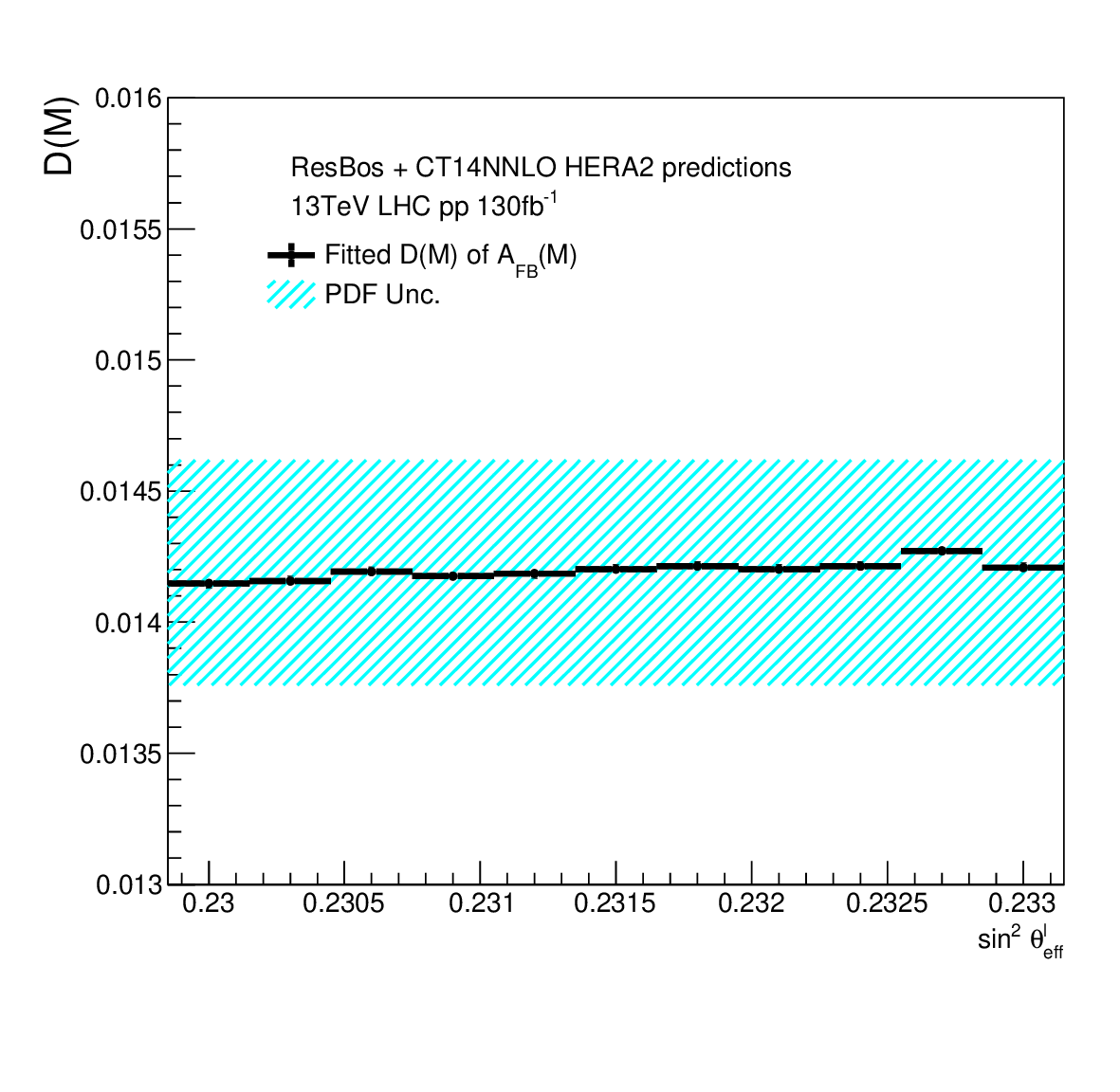}
\caption{\small The fitted $D(M)$ of $A_{FB}(M)$ in the $Z$-pole region of 80$<M<$100 GeV, as a function of $\effstw$. The error band corresponds 
to the PDF uncertainty of CT14HERA2, which is given at the $68\%$ C.L.}
\label{fig03:AFBslope}
\end{center}
\end{figure}

It can be seen that in the $Z$ pole region, the variation in the $A_{FB}(M)$ distribution with respect to different $\effstw$ inputs is approximately a global 
shift. This feature indicates that $\effstw$ governs more on the average $A_{FB}$ value around the $Z$ pole than the off-shell spectrum in the sideband region. 
To quantitatively describe this feature, we propose to express the $A_{FB}(M)$ spectrum in a factorization form of interpolation at the $Z$ mass pole:
\begin{eqnarray}\label{eq:slope}
A_{FB}(M) = D(M)\times(M - M_Z) + A_{FB}(M_Z)
\end{eqnarray}
where the parameter $D(M)$ represents the gradient of the $A_{FB}(M)$ shape and should be less dependent on $\effstw$, while the most $\effstw$-sensitive contribution 
in the $Z$-pole region is absorbed by the $A_{FB}(M_Z)$ term. Specifically, within the $Z$ pole region, the $A_{FB}(M)$ spectrum at the LHC is approximately a linear 
distribution and the $D(M)$ term is a constant slope. The linearity of the constant slope is depicted in Figure~\ref{fig02:AFBzpoleline}, with the input $\effstw$ set 
as 0.2345. Figure~\ref{fig03:AFBslope} shows the $\effstw$ dependence of the constant slope $D(M)$, obtained by fitting to the $A_{FB}(M)$ distribution in the $Z$ pole 
region, generated at various $\effstw$ values, with its PDF uncertainties predicted by the CT14HERA2 PDFs.
One can clearly see that within a large variation on the values of $\effstw$, the relative change on the slope is at a level of $\mathcal{O}(0.001)$, while the relative 
PDF-induced uncertainty on the slope is around $6\%$. Therefore, it is clear that the gradient of the $A_{FB}(M)$ spectrum is much more sensitive to PDF than to 
$\effstw$.

\begin{figure}[!hbt]
\begin{center}
\includegraphics[width=0.4\textwidth]{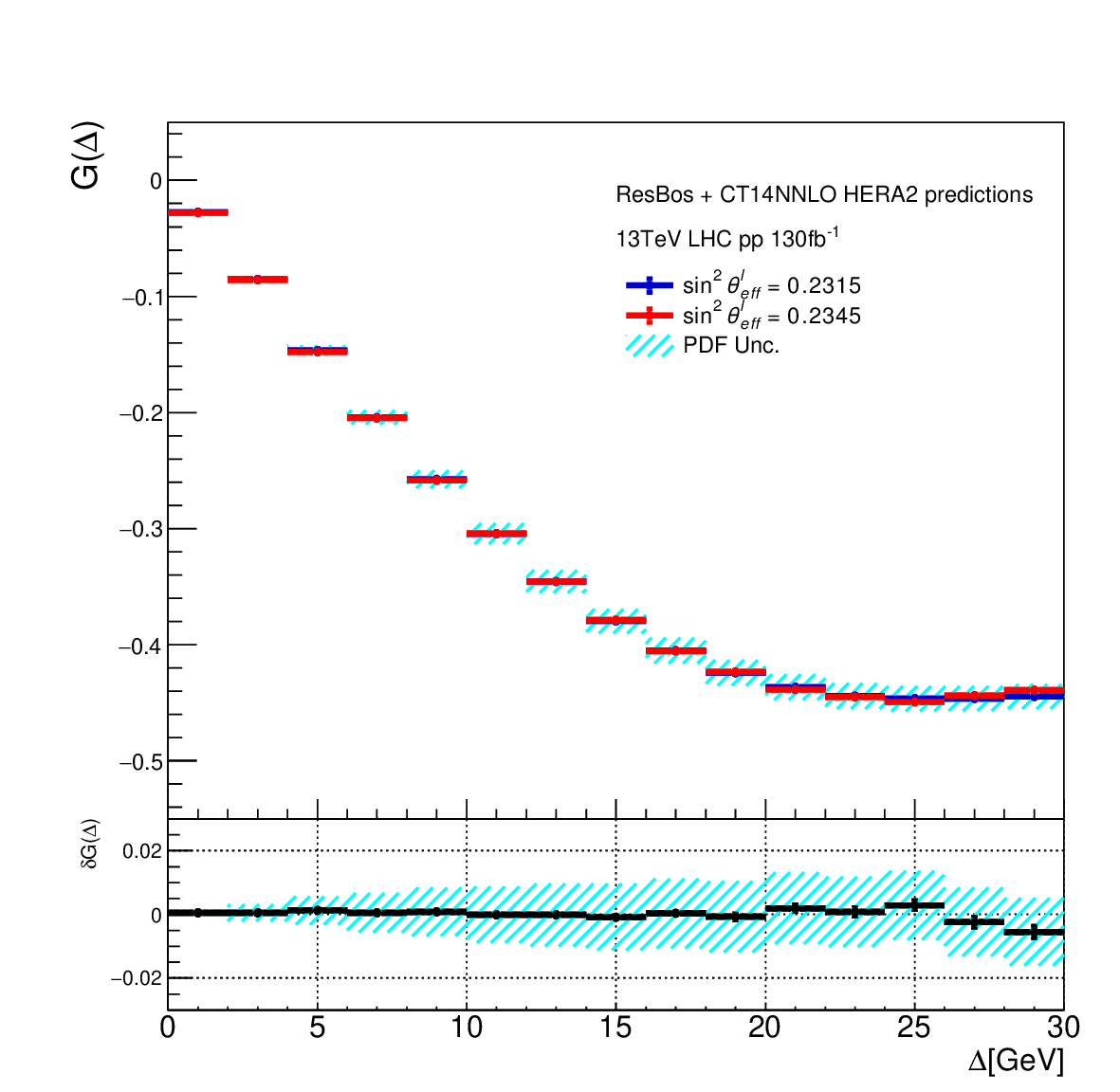}
\caption{\small The values of the observable $G(\Delta)$ with different $\effstw$ inputs of 0.2345 and 0.2315, respectively, shown as a function of $\Delta$ 
and calculated with averaged $A_{FB}$ in a 2 GeV mass bin. In the bottom panel, the solid lines depict the difference $\delta G(\Delta)$ of the two samples, 
while the error band corresponds to the PDF-induced uncertainty in $G(\Delta)$ calculated from CT14HERA2 PDFs.}
\label{fig04:ShapeObs}
\end{center}
\end{figure}

It should be noticed that Eq.~\ref{eq:slope} is only a simple example of defining the gradient of the $A_{FB}(M)$ spectrum to show the insensitivity of the shape information to $\effstw$. In fact, one could construct various experimental observables to represent the shape feature of the $A_{FB}(M)$ spectrum. Since 
the $A_{FB}(M)$ spectrum is not linear in the sideband region, we introduce a better observable, which needs no linear assumption, as:

\begin{eqnarray}
G(\Delta) = A_{FB}(90-\Delta) - A_{FB}(90+\Delta).
\label{eq04:ODelta}
\end{eqnarray}
This definition of shape observable reflects the relative difference of the $A_{FB}(M)$ values evaluated at the two mass points which are symmetrically located around 
90 GeV, close to the $Z$ mass pole. Figure~\ref{fig04:ShapeObs} shows the predicted values of the observable $G(\Delta)$ in two samples with different $\effstw$ 
inputs, respectively, as a function of the deviation $\Delta$ which is calculated with averaged $A_{FB}(90\pm\Delta)$ values in a typical 2 GeV mass bin 
in order to contain experimental conditions. As expected, the difference of the two samples in the whole symmetric mass deviation region, depicted as solid 
lines in the bottom panel, is negligibly small, manifesting that the new observable $G(\Delta)$ is insensitive to the $\effstw$ value. On the other hand, 
there are large PDF-induced uncertainties on $G(\Delta)$ when deviating from the $Z$ pole, indicating that this variable is sensitive to PDFs.

\section{Tests on Uncorrelated PDF Updating}

To numerically test how much the correlation between PDF and $\effstw$ could be reduced by the new variable, we perform a fast PDF updating by using the Error 
PDF Updating Method Package ({\sc ePump})~\cite{ePump-Schmidt:2018hvu, ePump-Hou:2019gfw} with pseudo-data samples. The {\sc ePump} tool provides a fast global 
fitting procedure consistent with that in the real PDF global fitting of CTEQ PDFs. To make this test comparable to previous studies~\cite{epumpAFBsideband}, 
we apply an ATLAS-like detector acceptance cut~\cite{ATLASdetector}. The central region leptons are defined with their pseudo-rapidity $|\eta|<2.5$, while 
the forward region leptons are defined with their pseudo-rapidity $2.5<|\eta|<4.9$. Leptons are further required to have their transverse momentum $p_T>25$ GeV. 
Only the Central-Central events (CC), where both leptons are in the central region, and the Central-Forward (CF) events, where one lepton in the central and 
the other in the forward region are used. The Forward-Forward (FF) events are removed due to the fact that these events are practically difficult to be precisely 
measured. For demonstration, the observed $G(\Delta)$ in the CC and CF categories are used as new pseudo-data input to {\sc ePump} for a fast PDF updating.

A set of nominal $G(\Delta)$ predictions given by CT14HERA2 error PDFs, used as theoretical inputs in the {\sc ePump} updating, are calculated from simulations 
generated with $\effstw=0.2315$. A pseudo-data sample is generated with $\effstw=0.2345$, of which the difference to that used in the nominal error set 
predictions is as large as 10 times the uncertainty on $\effstw$ from its best measurements~\cite{LEP-SLD, Tevatron-combine}. Since the error set predictions 
and the pseudo-data observations are derived from the same CT14HERA2 PDFs, an unbiased (or EW-uncorrelated) update should have no significant changes on any 
observable predictions, but have the PDF-induced uncertainty reduced. In this test, the correlation between PDF and $\effstw$ or the potential bias that might 
be induced into PDF updating is quantified by the deviation $\delta A_{FB}$, defined as the difference of the averaged $A_{FB}$ in the $Z$ pole region before 
and after the PDF updating. 

The numerical results of updating PDF using the new $G(\Delta)$ observable in the full mass region $60<M<120$ GeV, with CC, CF and CC+CF events, are listed in 
Table~\ref{tab01:ShapeUpdating}, respectively. It can be seen that the potential bias $\delta A_{FB}$ induced by the $G(\Delta)$ updating is negligibly small. 
For example, in the most sensitive CF category, the difference in the average value of $A_{FB}$, termed $\delta A_{FB}$, before and after updating the PDFs using 
the $G(\Delta)$ observable, is about $\sim 0.00003$, which is more than 20 folds smaller than the PDF-induced uncertainty ($\pm 0.00075$) of $A_{FB}$ itself after 
the updating, and also much smaller than its statistical fluctuation ($\pm 0.00017$).

\begin{table*}[hbt]
{\tiny
\begin{center}
\begin{tabular}{l|c|c|c}
\hline \hline
         & Updating using $CC$ events & Updating using $CF$ events & Updating using $CC$+$CF$ events \\
\hline
Average $A_{FB}$ in the $Z$-pole region of & & & \\
pseudo-data ($\effstw = 0.2345$) & & & \\ 
~~~~ in $CC$ events & 0.00714$\pm$0.00008(stat.) & 0.00714$\pm$0.00008(stat.) & 0.00714$\pm$0.00008(stat.) \\
~~~~ in $CF$ events & 0.03290$\pm$0.00017(stat.) & 0.03290$\pm$0.00017(stat.) & 0.03290$\pm$0.00017(stat.) \\
~~~~ in $CC+CF$ events & 0.01192$\pm$0.00007(stat.) & 0.01192$\pm$0.00007(stat.) & 0.01192$\pm$0.00007(stat.) \\
\hline
Theory prediction on average $A_{FB}$ & & & \\
 in the $Z$-pole region of $CC$ events ($\effstw = 0.2315$) & & & \\
~~~~ before update & 0.00873$\pm$0.00038(PDF) & 0.00873$\pm$0.00038(PDF) & 0.00873$\pm$0.00038(PDF) \\
~~~~ after update & 0.00874$\pm$0.00022(PDF) & 0.00874$\pm$0.00027(PDF) &  0.00873$\pm$0.00021(PDF) \\
~~~~ $\delta A_{FB}$[after - before] & 0.00001 & 0.00001 & 0.00000 \\
\hline
Theory prediction in $CF$ events, $\effstw = 0.2315$ & & & \\
~~~~ before update & 0.04220$\pm$0.00118(PDF) & 0.04220$\pm$0.00118(PDF)  & 0.04220$\pm$0.00118(PDF)  \\
~~~~ after update & 0.04220$\pm$0.00098(PDF) &  0.04223$\pm$0.00075(PDF) & 0.04220$\pm$0.00075(PDF) \\
~~~~ $\delta A_{FB}$[after - before] & 0.00000 & 0.00003 & 0.00000 \\
\hline
Theory prediction in $CC+CF$ events, $\effstw = 0.2315$ & & & \\
~~~~ before update & 0.01449$\pm$0.00053(PDF) &  0.01449$\pm$0.00053(PDF) &  0.01449$\pm$0.00053(PDF) \\
~~~~ after update & 0.01450$\pm$0.00035(PDF) & 0.01451$\pm$0.00035(PDF) &  0.01449$\pm$0.00031(PDF)\\
~~~~ $\delta A_{FB}$[after - before] & 0.00001 & 0.00002 & 0.00000 \\
\hline \hline
\end{tabular}
\caption{\scriptsize Average $A_{FB}$ in the $Z$-pole region of the pseudo-data and theory predictions. 
The PDF updating is performed using the full mass range $G(\Delta)$ in pseudo-data ($\effstw=0.2345$).
Statistical uncertainty corresponds to the data sample with an integrated luminosity of 130 fb$^{-1}$. }
\label{tab01:ShapeUpdating}
\end{center}
}
\end{table*} 

In Table~\ref{tab02:BiasChanges}, the potential bias after PDF updating by using different observables defined with the $A_{FB}(M)$ spectrum are compared. 
Again, one can see that the bias of the $G(\Delta)$ updating is nearly an order of magnitude less than that derived from an updating using 
the sideband-only $A_{FB}$, and two orders of magnitude less than that from an updating directly using the full mass range $A_{FB}$ spectrum.

\begin{table*}[hbt]
{\tiny
\begin{center}
\begin{tabular}{l|c|c|c}
\hline \hline
& Updating using full-mass $A_{FB}$ & Updating using sideband-only $A_{FB}$ & Updating using $G(\Delta)$ observable\\
& with $CF$ pseudo-data & with $CF$ pseudo-data & with $CF$ pseudo-data \\
\hline
Theory prediction on average $A_{FB}$ & & & \\
in $Z$-pole $CF$ events ($\effstw = 0.2315$) & & & \\
~~~~ before update & 0.04220 & 0.04220 & 0.04220 \\
~~~~ after update & 0.03889 & 0.04172 & 0.04223 \\
~~~~ $\delta A_{FB}$[after - before] & -0.00331 & -0.00048 & 0.00003 \\
\hline \hline
\end{tabular}
\caption{\scriptsize Average $A_{FB}$ in the $Z$-pole region of theory predictions of CF events. The PDF updating is performed using the full mass range $A_{FB}$, 
the sideband-only $A_{FB}$, and the $G(\Delta)$ with CF events in pseudo-data ($\effstw=0.2345$), respectively.}
\label{tab02:BiasChanges}
\end{center}
}
\end{table*}

\begin{figure}[!hbt]
\begin{center}
\includegraphics[width=0.30\textwidth]{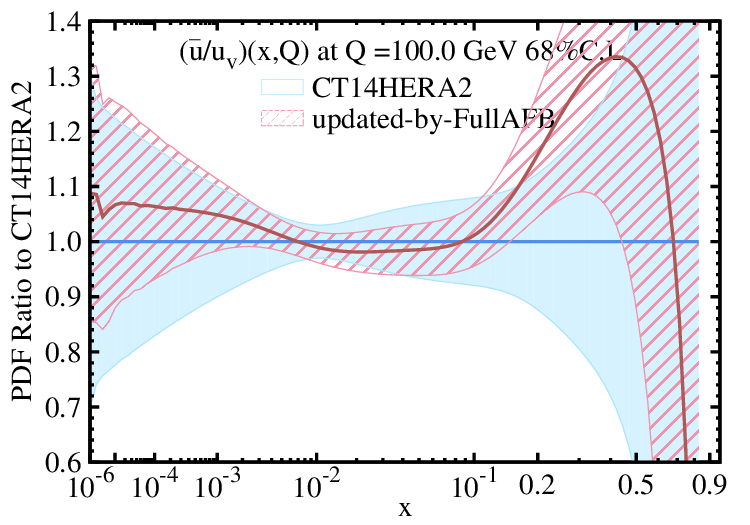}
\includegraphics[width=0.30\textwidth]{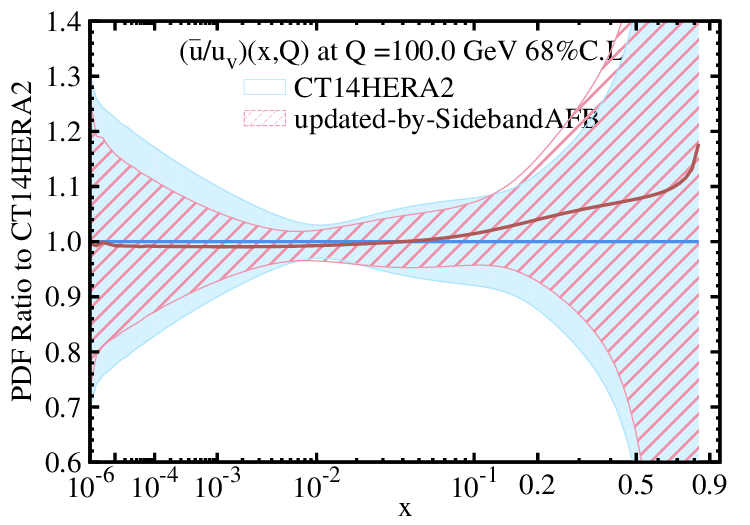}
\includegraphics[width=0.30\textwidth]{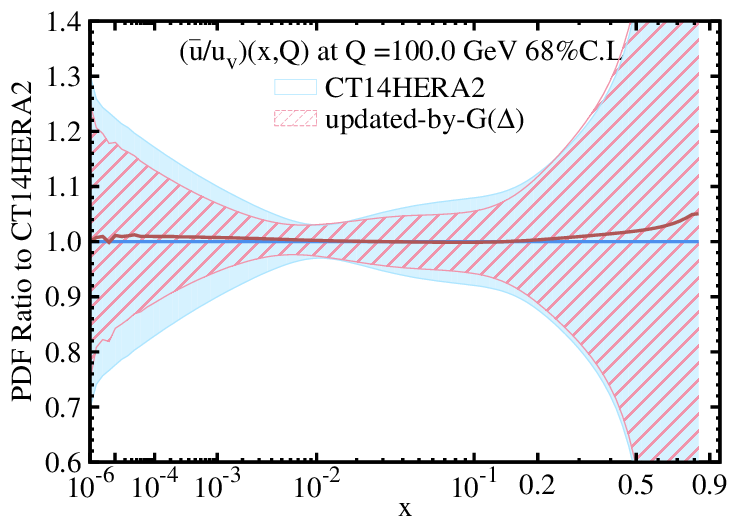}
\caption{\small 
The PDF ratio $\bar{u}/u_\text{valence}$ to the CT14HERA2 central prediction before and after updating PDF with $CF$ events in the 
pseudo-data ($\effstw=0.2345$). The left, middle and right plots correspond to using the full mass range $A_{FB}(M)$, 
the sideband $A_{FB}(M)$ and the full mass range $G(\Delta)$, respectively. The blue lines and error bands show the ratios before updating. 
The red lines and error bands show the ratios after updating. The $x$ axis is the fraction of the effective parton energy with respect 
to the proton.
}
\label{fig05:PDFuubar}
\end{center}
\end{figure}

\begin{figure}[!hbt]
\begin{center}
\includegraphics[width=0.30\textwidth]{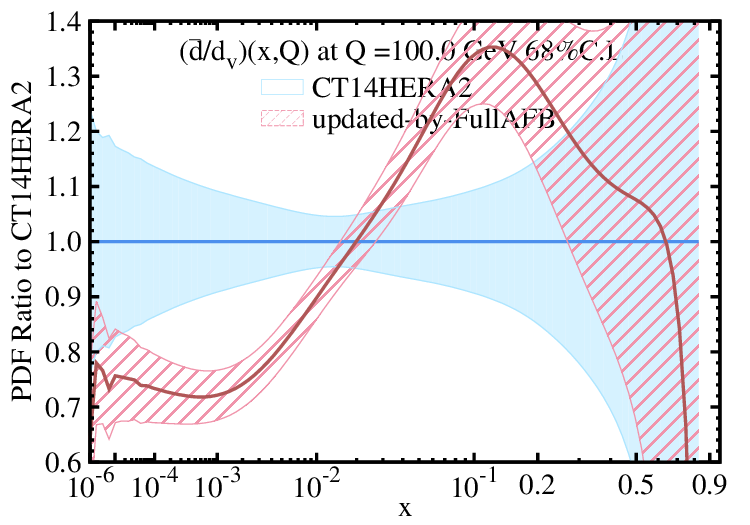}
\includegraphics[width=0.30\textwidth]{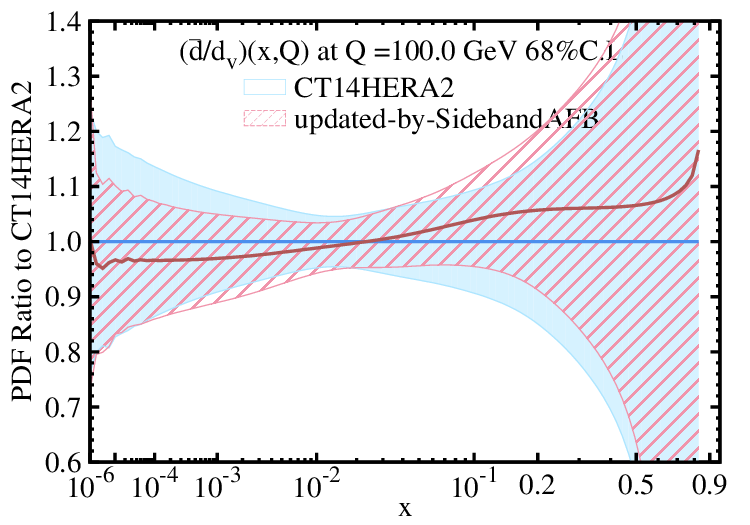}
\includegraphics[width=0.30\textwidth]{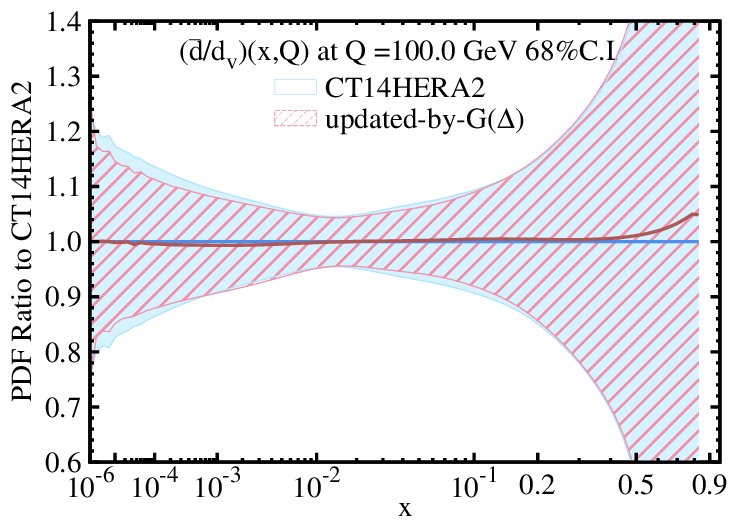}
\caption{\small 
Similar to	Fig.~\ref{fig05:PDFuubar}, but for 
the PDF ratio $\bar{d}/d_\text{valence}$.
}
\label{fig06:PDFddbar}
\end{center}
\end{figure}

To illustrate how much the correlation is reduced, we directly compare the PDFs before and after being updated with the full-mass region $A_{FB}$, the sideband-only 
$A_{FB}$, and the full-mass region $G(\Delta)$, as a function of $x$ in Figure~\ref{fig05:PDFuubar} and Figure~\ref{fig06:PDFddbar}, respectively. The ratio of the PDF 
predictions before and after updating are shown for the distributions of PDF ratios $\bar{u}/u_\text{valence}$ and $\bar{d}/d_\text{valence}$, as a function of $x$ at 
$Q=100$ GeV. A couple of important conclusions are noted below. First, the relative difference between the valance and sea PDFs of the first generation light quarks is 
sensitive to the $A_{FB}$ shape information. Second, the potential bias arising from the PDF-$\effstw$ correlation is significant when updating the PDFs by using 
the full mass range $A_{FB}$ spectrum, but would be reduced to negligible level by using the new $G(\Delta)$ observable.

Since the $G(\Delta)$ observable, or other similar observables describing the gradient of the $A_{FB}$ distribution, takes only part of the information from 
the $A_{FB}(M)$ spectrum, it is expected that the reduction on the PDF uncertainty will not be as large as that using the full mass range $A_{FB}(M)$ spectrum. 
However, the small difference in the efficiency of reducing the PDF uncertainties by the above two methods would diminish when a much larger data sample is 
collected at the future LHC experiments. Moreover, using the full mass range $A_{FB}(M)$ spectrum to update the PDFs will induce a much larger bias on 
the measurement of $\effstw$ than using the $G(\Delta)$ observable. Hence, it is advantageous to employ the proposed new observable $G(\Delta)$ in the analysis 
of precision electroweak data.

It is worth mentioning that measuring the shape information of $A_{FB}(M)$, such as $G(\Delta)$, would not require much more work after measuring the experimental observable $A_{FB}(M)$ at the LHC. The $G(\Delta)$ variable used in this study is just one of many possible observables. The principle is to remove the information 
of the overall average value of $A_{FB}(M)$, and use only the shape information of the spectrum. In this study, the pseudo-data has an input $\effstw$ value 
significantly different from those used in the theory predictions. It is expected that the experimental measurement of $\effstw$ will be further improved in 
the near future with higher precision, so the residual correlation associated with the shape information of the $A_{FB}(M)$ spectrum will become smaller than 
the numerical results presented in this article.

\section{Summary and discussion}

It has been a long-debated question about how to improve the accuracies of the PDF predictions and fundamental electroweak parameter determinations, 
with the precise W and Z boson measurements at hadron colliders. The key problem is that most of electroweak sensitive observables at hadron colliders 
are strongly correlated with PDF models, and therefore iterative fittings for the PDF and electroweak parameters will potentially induce bias into 
these two important physics sectors.

In this paper, a new method of reducing the correlation with the EW parameter $\effstw$ in the PDF updating is proposed, by using the forward-backward 
asymmetry $A_{FB}$ distribution of Drell-Yan $pp \rightarrow Z/\gamma^* \rightarrow \ell^+\ell^- +X$ processes. At the LHC, the $A_{FB}(M)$ spectrum as 
a function of the dilepton invariant mass, is found to be highly sensitive to the relative difference between parton distributions of the first generation 
quarks and antiquarks, thus can be used as an important input in the PDF global fits. Though the $A_{FB}(M)$ distribution at the LHC is determined by 
both the PDFs and the EW parameter altogether, it is concluded that the sensitivity of $A_{FB}$ on the $\effstw$ is governed by its overall average around 
the $Z$-pole region, while the shape (or gradient) of the $A_{FB}(M)$ spectrum is much less dependent on the $\effstw$ value but still highly sensitive 
to the PDF modeling. Accordingly, a well-defined experimental observable $G(\Delta)$ is introduced to represent the shape information of the $A_{FB}(M)$ spectrum, 
and is proposed to be used as an input data to the PDF updating. Numerical tests have been done and clearly demonstrate that the bias of directly using 
the $A_{FB}$ distribution in the PDF global fitting caused by it correlation with the electroweak physics can be significantly reduced, while the sensitivity 
to the PDF parameters is largely retained. 

This new method of using the gradient information of the $A_{FB}(M)$ spectrum in the PDF global fitting, and meanwhile using its average value around the $Z$-pole 
dominated by $A_{FB}(M_Z)$ to determine the $\effstw$, could provide a safe, uncorrelated and unbiased way for the high precision PDF update and electroweak 
measurement at the future high luminosity LHC experiments. The above discussion and proposal is based on the CT14HERA2 PDF predictions, and we have checked that 
similar conclusions also work for using the proposed $G(\Delta)$ to update other PDF sets, such as MMHT2014~\cite{MMHT2014}. In addition, although the quantative 
demonstration presented in the paper is mainly based on an ATLAS-liked experiment as an example, the idea and conclusions also hold for other hadron collider 
experiments such as CMS \cite{CMSdetector} and LHCb \cite{LHCbdetector}.

\begin{acknowledgments}
This work was supported by the National Natural Science Foundation of China under Grant No. 11721505, 11875245 and 12061141005.
This work was also supported by the U.S.~National Science Foundation
under Grant No.~PHY-2013791. C.-P.~Yuan is also grateful for the support from the Wu-Ki Tung endowed chair in particle physics.
\end{acknowledgments}


\begin{thebibliography}{99}
\vskip 0.25cm

\bibitem{CS-frame}
   J. C. Collins and D. E. Soper, Phys. Rev. D {\bf 16}, 2219 (1977). 

\bibitem{xFitterPaper1}
    E. Accomando, J. Fiaschi, F. Hautmann, and S. Moretti, Phys. Rev. D {\bf 99}, 079902 (2019).
   
\bibitem{xFitterPaper2}
    Accomando, E., Fiaschi, J., Hautmann, F. et al., Eur. Phys. J. C {\bf 78}, 663 (2018).
   
\bibitem{ArieMassConstrain}
    A. Bodek, J.-Y. Han, A. Khukhunaishvili, and W. Sakumoto, Eur. Phys. J. C {\bf 76}, 115 (2016). 
   
\bibitem{epumpAFBsideband}
    Y.~Fu, S.~Q.~Yang, M.~H.~Liu, L.~Han, T.~J.~Hou, C.~Schmidt, C.~Wang, C.~P.~Yuan, Chinese Physics C Vol. {\bf 45}, No. 5, 053001 (2021).

\bibitem{ATLAS-8TeV}
    ATLAS public note at https://atlas.web.cern.ch/Atlas/GROUPS/PHYSICS/CONFNOTES/ATLAS-CONF-2018-037/

\bibitem{CMS-8TeV}
    Sirunyan, A. M., Tumasyan, A., Adam, W. {\it et al.} (CMS Collaboration), Eur. Phys. J. C {\bf 78}, 701 (2018).

\bibitem{LHCb-8TeV}
    R.~Aaij {\it et al.} (LHCb Collaboration), JHEP {\bf 11}, 190 (2015).

\bibitem{LEP-SLD}
    G. Abbiendi {\it et al.} (LEP Collaborations ALEPH, DELPHI, L3, and OPAL; SLD Collaboration; LEP Electronweak Working Group; 
    SLD Electroweak and Heavy Flavor Groups), Phys. Rep. {\bf 427}, 257 (2006).

\bibitem{Tevatron-combine}
    T. Aaltonen {\it et al.} (CDF and D0 Collaborations), Phys. Rev. D {\bf 97}, 112007 (2018).

\bibitem{Dulat:2015mca}
    S.~Dulat, T.~J.~Hou, J.~Gao, M.~Guzzi, J.~Huston, P.~Nadolsky, J.~Pumplin, C.~Schmidt, D.~Stump and C.~Yuan,
    Phys. Rev. D {\bf 93}, no.3, 033006 (2016).

\bibitem{Hou:2016nqm}
    T.~J.~Hou, S.~Dulat, J.~Gao, M.~Guzzi, J.~Huston, P.~Nadolsky, J.~Pumplin, C.~Schmidt, D.~Stump and C.~P.~Yuan,
    Phys. Rev. D {\bf 95}, no.3, 034003 (2017).

\bibitem{Hou:2019gfw}
    T.~J.~Hou, Z.~Yu, S.~Dulat, C.~Schmidt and C.~P.~Yuan,
    Phys. Rev. D {\bf 100}, no.11, 114024 (2019).

\bibitem{Hou:2019efy}
    T.~J.~Hou, J.~Gao, T.~J.~Hobbs, K.~Xie, S.~Dulat, M.~Guzzi, J.~Huston, P.~Nadolsky, J.~Pumplin and C.~Schmidt, \textit{et al.}
    Phys. Rev. D {\bf 103}, no.1, 014013 (2021).

\bibitem{ParticleDataGroup:2018ovx}
    M.~Tanabashi \textit{et al.} [Particle Data Group],
    Phys. Rev. D {\bf 98}, no.3, 030001 (2018).

\bibitem{Lai:2010nw}
    H.~L.~Lai, J.~Huston, Z.~Li, P.~Nadolsky, J.~Pumplin, D.~Stump and C.~P.~Yuan,
    Phys. Rev. D {\bf 82}, 054021 (2010).

\bibitem{RESBOS}
    C. Bal\'{a}zs, C.-P. Yuan, Phys. Rev. D {\bf 56}, 5558 (1997).

\bibitem{ePump-Schmidt:2018hvu}
    C.~Schmidt, J.~Pumplin, and C.~P.~Yuan, Phys. Rev. D {\bf 98}, no.9, 094005 (2018).

\bibitem{ePump-Hou:2019gfw}
    T.~J.~Hou, Z.~Yu, S.~Dulat, C.~Schmidt and C.~P.~Yuan, Phys. Rev. D {\bf 100}, no.11, 114024 (2019).

\bibitem{ATLASdetector}
  G. Aad {\it et al.} (ATLAS Collaboration), JINST {\bf 3}, S08003 (2008).

\bibitem{MMHT2014}
  Harland-Lang, L.A., Martin, A.D., Motylinski, P. {\it et al.}, Eur. Phys. J. C {\bf 75}, 204 (2015).

\bibitem{CMSdetector}
  S. Chatrchyan {\it et al.} (CMS Collaboration), JINST {\bf 3}, S08004 (2008). 

\bibitem{LHCbdetector}
  A. A. Alves Jr. {\it et al.} (LHCb collaboration), JINST {\bf 3}, S08005 (2008).
	
\end{thebibliography}
\end{document}